\documentstyle[sprocl,psfig]{article}

\def\Journal#1#2#3#4{{#1} {\bf #2}, #3 (#4)}

\def\AJ{\em Astron. Journal}
\def\ApJ{\em Astroph. Journal}
\def\AA{\em Astron. and Astroph.}
\def\Na{\em Nature}

\def\be{\begin{equation}}
\def\ee{\end{equation}}
\def\bea{\begin{eqnarray}}
\def\eea{\end{eqnarray}}

\begin{document}

\title{AN X-RAY TEMPERATURE MAP OF COMA}

\author{ULRICH G. BRIEL}

\address{Max-Planck-Institut f\"ur extraterrestrische Physik,\\
85740 Garching, Germany\\ E-mail: ugb@mpe-garching.mpg.de} 

\author{J. PATRICK HENRY}

\address{Institute for Astronomy, University of Hawaii,\\
Honolulu, HI 96822, USA \\E-mail: henry@uhifa.ifa.hawaii.edu}

\maketitle\abstracts{
We present an X-ray temperature map of the Coma cluster of
galaxies obtained with the ROSAT PSPC. As expected from the X-ray
surface brightness distribution the intracluster gas of Coma is 
not isothermal. The temperature structure resembles a bow shock of 
hot gas produced by the passage of the subcluster around NGC 4839 
through the main cluster, confirming hydrodynamical simulations.
}

\section{Introduction}

\noindent 
The Coma cluster of galaxies was long considered to be the archtype
of a relaxed virialized cluster in a state of dynamical equilibrium
(c.f. Kent and Gunn~\cite{kg}). There were however conflicting claims 
that Coma shows substructure, seen as a clumping of galaxies around 
the brightest galaxies in the cluster (c.f. Baier~\cite{ba}, Fitchett
and Webster~\cite{fw}, and Mellier {\it et al.}~\cite{me}). On the 
other hand, Geller and Beers~\cite{gb} and Dressler and Schectman~\cite{ds}
claim that there is no statistically significant 
structure in the Coma cluster. The first unequivocal evidence for 
substructure in Coma came from an X-ray image, obtained during the all
sky survey of the ROSAT satellite (Briel, Henry and B\"ohringer~\cite{bhb}).
They found diffuse X-ray emission from the regions of the NGC 4839 and 
4911 subgroups and interpreted the 4839 group to be in the process of 
merging with the main cluster. Long ROSAT PSPC pointed observations revealed
even more irregular cluster structure and X-ray emission from a number
of bright galaxies (White, Briel and Henry~\cite{wbh} and Dow and 
White~\cite{dw}). Applying the wavelet transform analysis to these pointed
observations, more significant substructure was found in the core of Coma 
(c.f. Biviano {\it et al.}~\cite{bdg} and Vikhlinin, Forman and 
Jones~\cite{vfj}). Using these observational results, Burns 
{\it et al.}~\cite{brl} made hydrodynamic/N-body simulations and concluded
that the 4839 group has already passed through the Coma cluster. More evidence
for the merging scenario came from the first temperature
map of the intracluster gas of Coma, obtained from the ASCA observation
(Honda {\it et al.}~\cite{hhw}) and from further simulations by Ishizaka and 
Mineshige~\cite{im}.

In this paper we report on a more detailed temperature map of Coma, obtained 
from the pointed ROSAT PSPC observations. Although ROSAT only observes in an
energy band from 0.2 to 2.5 keV, we have shown on several clusters of galaxies
that it is possibile to determine the usual high cluster temperatures, given
a sufficiently high photon statistic (c.f. Briel and Henry~\cite{bh1} and
Henry and Briel~\cite{hb1}).
 
\begin{table}[t]
\caption{Journal of Observations. \label{tab:exp}}
\vspace{0.2cm}
\begin{center}
\footnotesize
\begin{tabular}{|c|c|c|c|}
\hline
 & RA & DEC & Exposure  \\
 & (2000) & (2000) & (ksec)  \\
Date & h  min sec & $^o$ \ ' \ '' & MV $\le$ 170 \\
\hline
1991 Jun 16 \hspace{1.15cm} & 12 57 43.20 & +27 36 00.0 & 20.3 \\
1991 Jun 16 -- Jun 17 & 12 59 45.60 & +27 48 00.0 & 21.5 \\
1991 Jun 17 -- Jun 18 & 12 59 45.60 & +27 58 12.0 & 20.8 \\ 
1991 Jun 18 -- Jun 19 & 13 00 31.20 & +28 07 48.0 & 21.2 \\
\hline
\end{tabular}
\end{center}
\end{table}

\begin{figure}[t]
\begin{minipage}[t]{9.8cm}
\psfig{figure=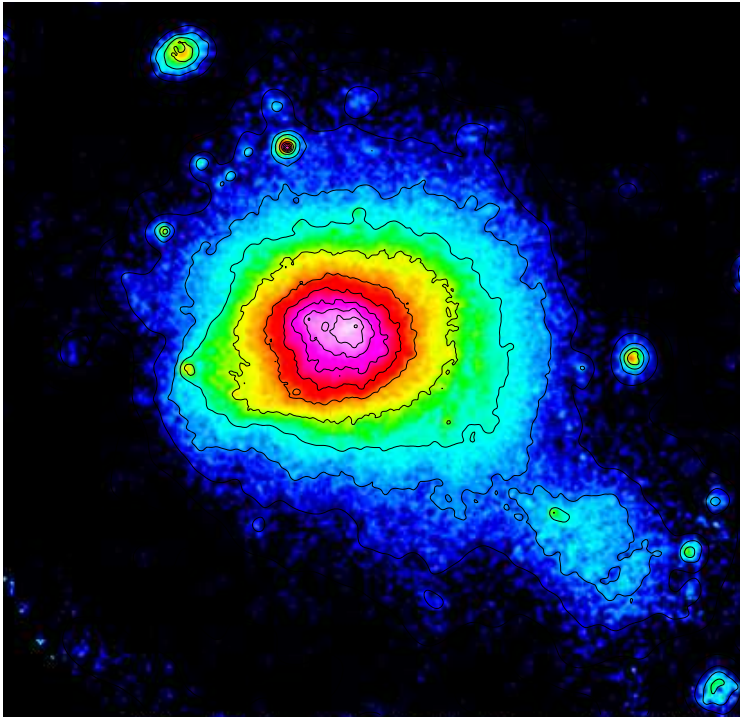,height=9.5truecm,angle=0,clip=}
\caption{False color image of the surface brightness distribution of Coma in 
the 0.5 -- 2.4 keV band with overlain contours. The contour levels are 
0.125,0.25,0.5,1,2,4,8,12,16,20,24,28$\times$10$^{-4}$counts sec$^{-1}$
(16 arcsec$\times$16 arcsec)$^{-1}$, including a background of $\approx$ 0.1 
in the same units. (Some of the lower contours are not visible).
\label{fig:surf}}
\end{minipage}
\end{figure}

\begin{figure}[t]
\begin{minipage}[t]{10.5cm}
\psfig{figure=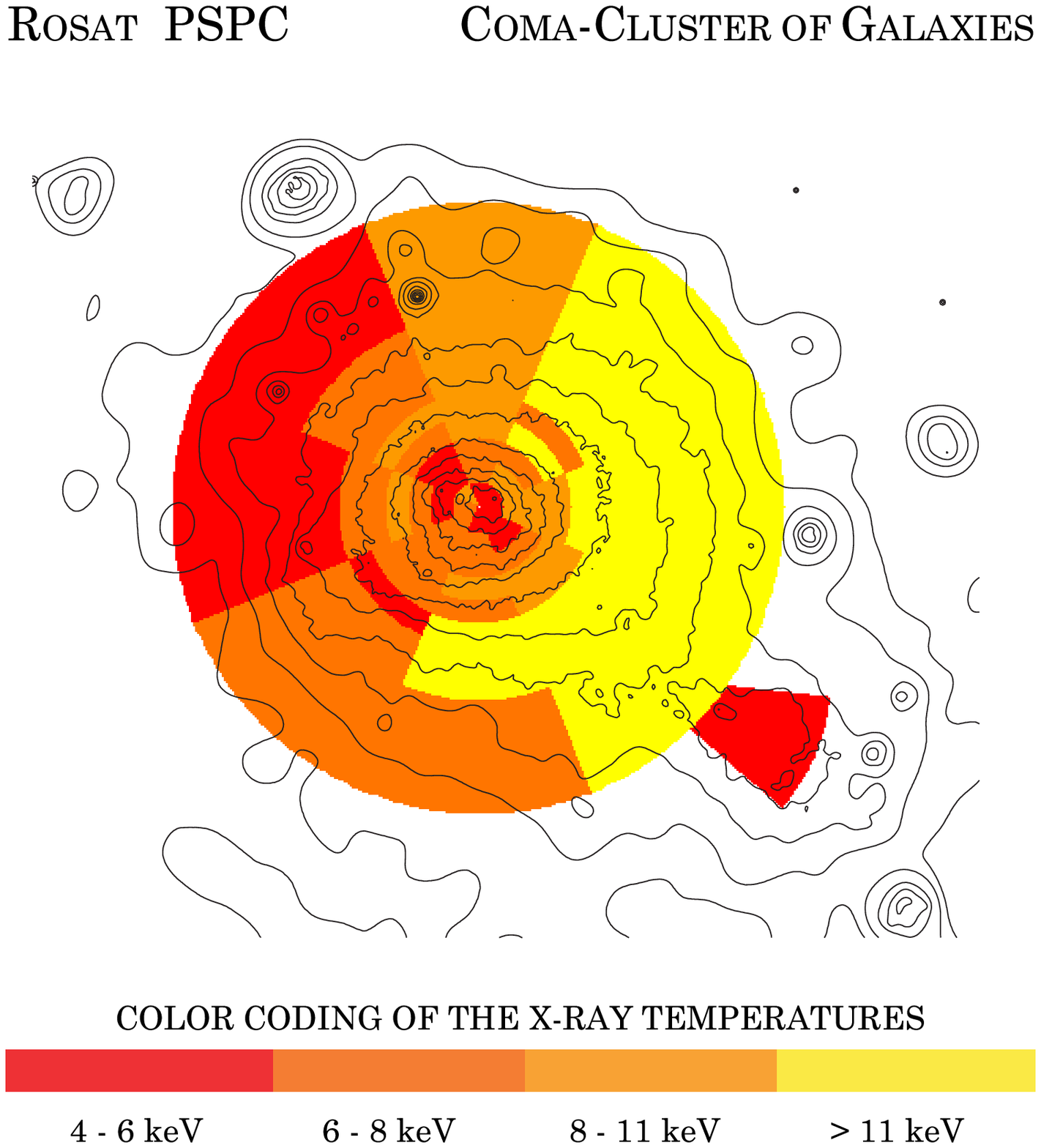,height=10.5truecm,angle=0,clip=}
\caption{Color--coded temperature map of Coma with superposed surface 
brightness contours in the 0.5 -- 2.4 keV band. The contour levels are 
the same as in Figure 1.
\label{fig:temp}}
\end{minipage}
\end{figure}

\section{Observations and Data Reduction}

The Coma cluster was in the field of view of 4 pointed observations performed
with the PSPC (Pfeffermann {\it et al.}~\cite{pbh}) on board the ROSAT 
satellite (Tr\"umper~\cite{tru}). 
In Table 1 we show the journal of the observations with the pointing 
directions and the accepted on axis exposure times. 
A spatial analysis was done by White, Briel and Henry~\cite{wbh} from
which we show in Figure 1 the surface brightness distribution in the 0.5 -- 2.4 
keV energy band. To emphasize faint structures they used an increasingly 
large smoothing at increasing lower surface brightness. For more details
on the procedure and on the interpretation of the image see references 8 and 9.

To obtain a temperature map of the intra cluster gas, we used essentially
the same procedure we have described in detail in the paper about the 
temperature map of A2142 (Henry and Briel~\cite{hb2}). The main difference 
was that for Coma we used the Rev2 data of the SASS, using the EXSAS command
{\it process/ct} to correct for spatial variations of the PSPC gain and to 
adjust the overall gain of the PSPC within 33 arcmin diameter to a value 
consistent with the temperature of 8.11 $\pm$ 0.04 keV, as measured with the 
GINGA satellite (David {\it et al.}~\cite{dsj}). The neccessay adjustment of 
the gain was in the order of $\pm$1\% for the four observations. 
Background-subtracted spectra from sectors of rings centered near the center 
of Coma from each pointing were summed, after correcting each photon for 
vignetting, and then fitted to Raymond--Smith models with the heavy-element
abundance fixed to 0.22 of their solar values. 

\noindent As usual, photons from point sources were excluded during all 
spectral fits. Special care was taken to obtain the background level outside 
at least 50 arc\-min from the cluster center, at which Coma shows a surface 
brightness of less then 0.3\% from its peak brightness, which is less 
than 20\% of the background. In Figure 2 we show the result of the spectral 
fitting of the different regions as a color representation of the temperatures 
with overlain contour map of the surface brightness from Figure 1. The four
temperature bands are roughly separated by $\approx$ 1 $\sigma$ (for one 
parameter of interest). There are three main points to note: (1) The region
around NGC 4839 shows a low temperature of 4.8 +1.1/-0.8 keV, consistent
with the typical temperature of a group of galaxies. (2) There is a hot 
arc-shaped region at the west side of the cluster where the NGC 4839 group is 
located, confir\-ming  the result of the ASCA measurement~\cite{hhw}.
(3) Within 30 arcmin diameter, where the cluster has its highest surface 
brightness, we find significant temperature structure on scales of a few arcmin
(1 arcmin corresponds to 40 h$_{50}^{-1}$ kpc).


\section{Conclusions}

The intracluster gas of the Coma cluster of galaxies shows significant 
temperature structure on large scales and down to a few arcmin scale. The
large scale temperature structure resembles an arc-shaped bow shock at a 
significant higher temperature compared with the rest of the cluster, located 
at the same side where the cooler galaxy group around NGC 4839 is found. This 
bow shock can be interpreted as the result of the passage of the group 
through the main cluster, as it was suggested by hydrodynamic/N-body
simulations of merger events~{\cite{brl}$^,$~{\cite{im}. Hence, this temperature
map of the Coma cluster of galaxies might have answered the open question
whether the subgroup is on its way through the cluster or if it already has 
passed the cluster core. If the answer is in fact in favor of the post-merger
scenario, then an other puzzle occurs: where does the intragroup gas come from?
Was there enough time after the passing to built up new gas, or was the 
stripping of the group gas while passing through the cluster inefficient? More 
simulations of the merging/passing of a group with/through a cluster are needed 
to clarify those questions.

{\small

\section*{Acknowledgements}
The ROSAT project is supported by the BMBF. JPH thanks Prof. J. Tr\"umper and 
the ROSAT group for their hospitality during the course of this research.  
JPH was supported by NASA grant NAG5-1789, and NATO grant CRG 910415.
UGB and JPH thank Simon White for using his Coma Rev2-data prior to the release
into the public archive.

}
\end{document}